\documentclass[aps,pra,preprintnumbers,amsmath,amssymb,footinbib]{revtex4}
\usepackage{graphicx}
\usepackage{bm}
\usepackage{dcolumn}
\usepackage[breaklinks=true,colorlinks,citecolor=black,linkcolor=blue,urlcolor=blue]{hyperref}

\begin{document}

\title{XUV frequency comb operation in an astigmatism-compensated enhancement cavity}

\author{J. Nauta,$^{1,2}$}\email{janko.nauta [at] mpi-hd.mpg.de}
\author{J.-H. Oelmann$^{1,2}$}
\author{A. Borodin$^{1}$}
\author{A. Ackermann$^{1}$}
\author{P. Knauer$^{1}$}
\author{I.~S.~Muhammad$^{1}$}
\author{R. Pappenberger$^{1}$}
\author{T. Pfeifer$^{1}$}\email{thomas.pfeifer [at] mpi-hd.mpg.de}
\author{J.~R.~Crespo López-Urrutia$^{1}$}\email{crespojr [at] mpi-hd.mpg.de}
\affiliation{$^1$Max-Planck-Institut für Kernphysik, Saupfercheckweg 1, 69117 Heidelberg, Germany}
\affiliation{$^2$Heidelberg Graduate School for Physics, Ruprecht-Karls-Universität Heidelberg, Im Neuenheimer Feld 226, 69120 Heidelberg, Germany}

\date{\today}

\begin{abstract}
We have developed an extreme ultraviolet (XUV) frequency comb for performing ultra-high precision spectroscopy on the many XUV transitions found in highly charged ions (HCI). Femtosecond pulses from a 100\,MHz phase-stabilized near-infrared frequency comb are amplified and then fed into a femtosecond enhancement cavity (fsEC) inside an ultra-high vacuum chamber. The low-dispersion fsEC coherently superposes several hundred incident pulses and, with a single cylindrical optical element, fully compensates astigmatism at the $w_0=15$\,$\mu$m waist cavity focus. With a gas jet installed there, intensities reaching $\sim10^{14}$\,W/cm$^2$ generate coherent high harmonics with a comb spectrum at 100\,MHz rate. We couple out of the fsEC harmonics from the 7th up to the 35th (42\,eV; 30\,nm) to be used in upcoming experiments on HCI frequency metrology.
\end{abstract}

\maketitle

\section{Introduction}
Soon after its invention, the frequency comb (FC) revolutionized the field of optical frequency metrology \cite{Udem2002}. The sharp modes of a phase-coherent pulse train with stabilized offset frequency $f_{\mathrm{CEO}}$ and repetition frequency $f_{\mathrm{rep}}$ act as a ruler in frequency space and are given by
\begin{equation}
f_{\mathrm{n}}=f_{\mathrm{CEO}}+n f_{\mathrm{rep}},
\label{eq:FC_comb_intro}
\end{equation}
where the mode index $n$ is a large ($\sim 10^6$) integer. These modes form a comb structure that enables absolute frequency determinations with unprecedented precision throughout the infrared and optical regions. The FC has become an indispensable tool for optical clocks \cite{Ludlow2015}, and found novel applications, e.~g., in the fields of molecular spectroscopy \cite{Picque2019}, exoplanet discovery \cite{Wilken2012}, attosecond science \cite{Baltuska2003} and optical communications \cite{Torres-Company2013}. In 2005, the FC spectral range was extended to the extreme ultraviolet (XUV) region below 100\,nm using the process of high-harmonic generation (HHG) to produce odd harmonics of the central wavelength of the original comb \cite{Jones2005,Gohle2005} displaying the same mode spacing. This technique has rapidly evolved since its introduction; the generated powers have increased by roughly six orders of magnitude and are currently on the order of mW per harmonic \cite{Porat2018}; the highest photon energies have surpassed 100\,eV \cite{Carstens2016}; and it was shown that such combs can generate radiation with coherence times exceeding $1$\,s \cite{Benko2014}. 

Exploiting well-developed optical and near infrared (NIR) laser-stabilization techniques, XUV combs can directly excite atomic transitions in the XUV \cite{Cingoz2012}. A very interesting proposal is using an XUV comb to investigate the 1S-2S transition in He$^+$ at 60\,nm, because it provides a benchmark for calculations of bound-state quantum electrodynamics, and fundamental quantities like the Rydberg constant can be extracted from its value \cite{Dreissen2019}. Another enormously interesting case for XUV combs is exciting the $8.2$\,eV isomeric state of the $^{229}$Th nucleus \cite{Wense2016,Seiferle2019a}, currently object of many investigations. With a wavelength of $149.7 \pm 3.1$\,nm, this unique nuclear transition is a promising candidate for a proposed clock of extraordinary stability \cite{Wense2018,Peik2003,Campbell2012,Wense2020} due to the long lifetime of the isomeric state \cite{Seiferle2017}, its extreme insensitivity to external perturbations and its sensitivity as a probe of new physics beyond the Standard Model \cite{Flambaum2006,Seiferle2019}. 

Finally, many recent theoretical studies have confirmed that highly charged ions (HCI) are also very promising systems for a new generation of atomic clocks also due to their insensitivity to perturbations and extreme metastability \cite{Berengut2012a,Dzuba2012,Kozlov2018b}; experiments have made them available \cite{Schmoeger2015} and demonstrated their feasibility \cite{Micke2020}. In the XUV and x-ray regions, neutrals become photoionized, while HCI remain stable under irradiation with free-electron lasers \cite{Epp2007,Bernitt2012} and synchrotrons \cite{Rudolph2013}, enabling fluorescence spectroscopy on dipole-allowed transitions at those high photon energies. At the same time, many suitably narrow clock transitions can be found in HCI by selecting an appropriate charge state, atomic number and isotope, and certain isoelectronic sequences also serve as fine probes of new physics \cite{Berengut2010,Berengut2012}. The feasibility of our proposed XUV experiments \cite{Crespo2016,Nauta2017,Oelmann2019} with HCI has just been investigated theoretically \cite{Lyu2020}.

After the demonstration of coherent laser spectroscopy of HCI \cite{Micke2020} in the optical range, we have developed a new XUV frequency comb \cite{Nauta2017} for extending that technique into the XUV region. In order to perform high-accuracy frequency determinations in HCI, the comb-line spacing needs to be much larger than the linewidth of the targeted transition. A fiber-based, NIR comb laser operates with a repetition rate of 100\,MHz. At such values, it is far more difficult to reach the peak powers required for HHG $>10^{13}$\,W/cm$^2$ than with conventional kHz laser systems. Therefore, we additionally raise the power of the femtosecond NIR pulses in a passive femtosecond enhancement cavity (fsEC) installed in a large vacuum chamber to suppress XUV absorption. Within the cavity, which has an optical path length matching the NIR pulse separation, the incident NIR pulses are coherently overlapped and strongly focused on a gas target. The light-matter interaction in the cavity focus produces high harmonics that propagate co-linearly with the NIR pulse. In our intended use, they will be coupled out of the cavity and guided to a superconducting Paul trap storing cold HCI \cite{Stark2020}. 

The fsEC is the crucial element of the XUV comb, and the size and shape of the NIR focus it produces defines the interaction volume where HHG takes place. To maximize the laser intensity and generate a symmetric XUV beam, a round focus with a similar waist size in the sagittal (vertical) and tangential (horizontal) plane is desirable. A standard bow-tie cavity geometry, however, inherently produces an asymmetrical focal volume, arising from non-normal incidence on the concave focusing mirrors. To reduce this astigmatism, a non-planar geometry can be used \cite{Winkler2016}. In such a configuration, the larger number of mirrors leads to additional NIR dispersion and scattering losses, although the sagittal and tangential foci still do not overlap. In a planar bow-tie configuration, astigmatism can be avoided by using parabolic focusing mirrors \cite{Dupraz2015} or two identical cylindrical cavity mirrors \cite{Carstens2013}. Although the former solution is elegant, parabolic mirrors are costly, and cavity alignment becomes more difficult as the rotation of each parabolic mirror is an additional degree of freedom. The latter method works well and does not have these disadvantages, but requires a symmetric position of the mirrors around the focus. A fully astigmatism-compensated fsEC with an ideal asymmetry for reflective output coupling of the XUV radiation has not yet been realized.

In this work, we present a novel fsEC design with a single cylindrical input coupler (IC) optic that entirely compensates astigmatism near the focal volume. It consists of just five optical elements, thus minimizing dispersion and scattering losses as well as risks of mirror degradation due to carbon cracking \cite{Hollenshead2006}. Between the focus and one of the focusing mirrors, a grating-mirror \cite{Yost2008} deflects the generated XUV radiation out of the fsEC while reflecting the NIR pulses in zeroth order. This arrangement leaves enough space around the interaction volume for the differential pumping system, which is in turn required to handle the dense gas jet \cite{Nauta2020a}.

\section{Cavity design considerations}
In principle, it is desirable for a large HHG yield to maximize the cavity focal volume \cite{Cingoz2012}, but increasing its size also reduces the NIR intensity and dramatically decreases the yield. A large focal volume thus requires a high cavity finesse, or a very intense laser. In order to reduce the sensitivity of the cavity mode to dispersive effects caused by the target gas plasma, it is preferable to keep the finesse low, $\ll1000$ \cite{Allison2011}. Incident laser powers above $\sim$100\,W require careful control of thermal effects in the amplifier fibers and are technically challenging. Shortening the injecting NIR pulses to increase their peak intensity is limited by the low-dispersion bandwidth of the high-reflectivity cavity mirrors. Finally, a small focus size is advantageous to reduce the amount of plasma persisting between pulses in the focal region, since ions leave it faster \cite{Saule2018}. Thus, for optimal HHG conditions using a low finesse cavity and a limited amount of incident laser power, strong focusing is preferred. 

For direct frequency-comb spectroscopy, only a single mode needs to interact with the HCI. The grating-mirror (GM) separates the different harmonic orders and coarsely selects the wavelength to be send to the ion. Choosing the GM as XUV output coupler \cite{Yost2008} has the great advantage over other methods, such as using Brewster plates \cite{Jones2005,Gohle2005} or geometrical outcoupling \cite{Pupeza2013,Zhang2020a}, that it both separates the XUV light from the NIR light and spatially disperses the harmonic orders with a single reflection. The GM interacts with the NIR cavity beam only at the surface, such that losses and dispersion can be minimized by choosing a suitable coating. Furthermore, the GM can efficiently diffract a broad wavelength bandwidth, such that a large variety of transitions can be probed in many different HCI species. Finally, the GM has been demonstrated to function over an extended time period without severe degradation effects \cite{Mills2019a}. 

Recent experiments have shown that feeding the target jet with high-pressure gas speeds up its expansion and reduces plasma effects in the focal region. This significantly improves the HHG yield \cite{Porat2018}, but requires stronger pumping, which we improve by means of a sophisticated multi-stage differential pump system around the nozzle. This structure excludes cavity beams and optical elements within 10\,mm of the focus region. 

\begin{figure}[tb]
\centering\includegraphics[width=\textwidth]{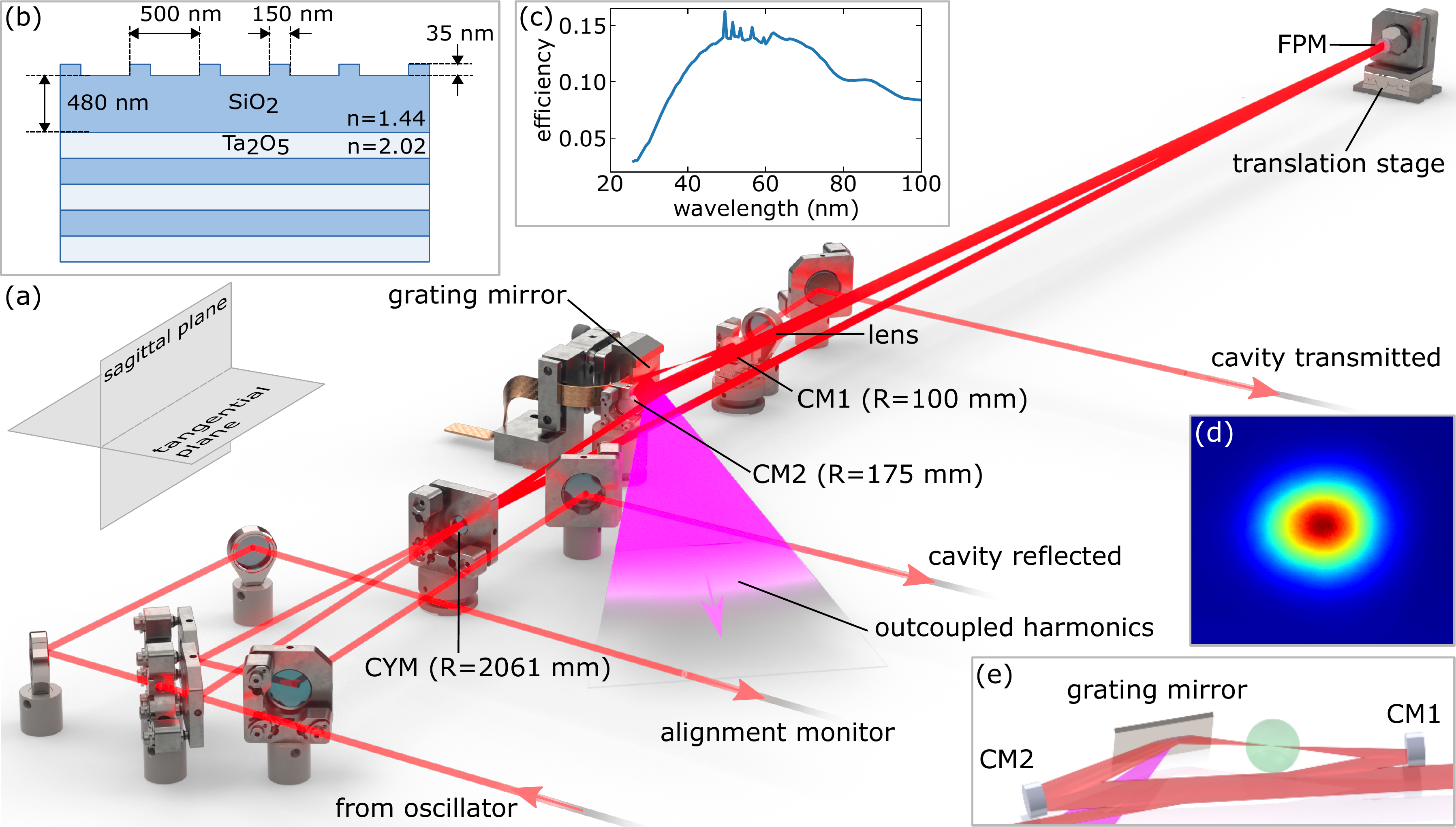}
\caption{(a) Scheme of the in-vacuo elements of the enhancement cavity. After reflection by a single mirror, the input beam enters the cavity through the cylindrical input coupler (CYM). Its alignment is monitored using the weak transmitted beam, which is guided out of the chamber. The beam reflected from the input side of the CYM is used for locking the cavity length by correcting it with the piezo-driven flat cavity mirror (FPM) applying the Pound-Drever-Hall (PDH) method. The short cavity arm is formed by two curved mirrors (CM1, CM2) and the grating mirror. For monitoring the intra-cavity mode profile and circulating power we use the small leakage through CM1. An arrow indicates the propagation direction of the harmonics. (b) Principle of the grating mirror (GM). It consists of a quarter-wave dielectric stack with a thick top layer with an etched grating structure with a short period, reflecting NIR but diffracting XUV radiation. (c) Calculated diffraction efficiency of the GM for the wavelengths of interest. (d) Profile of the Gaussian cavity mode from the cavity-transmitted beam at a distance of 0.78\,m after the mirror as recorded with a CCD camera. (e) Close-up of the focus region. The free space with a radius of 10\,mm around the cavity focus is marked by a green sphere. }
\label{fig:cav_schem}
\end{figure}

We design the geometry of the cavity with the ABCD matrix formalism, where the propagation of a Gaussian beam is described by a complex $q$ parameter,
\begin{equation}
\frac{1}{q(z)}=\frac{1}{R(z)}-i \frac{\lambda}{\pi w(z)^{2}}.
\label{eq:q_param}
\end{equation}
$R(z)$ is the wavefront radius of curvature and $w(z)$ and the beam waist size, which both depend on the longitudinal position $z$. The beam propagation through an optical system is described with a ray-transfer matrix defined as
\begin{equation}
M_{\mathrm{tot}}=\left(\begin{array}{cc}
A & B \\
C & D
\end{array}\right).
\end{equation}
Using matrices for common optical elements from the literature \cite{Siegman1986}, the total transfer matrix of the system results from their matrix product. The $q$ parameter at the exit evolves from the incident beam $q_1$ according to
\begin{equation}
q_{2}=\frac{A q_{1}+B}{C q_{1}+D}.
\label{eq:q_prop}
\end{equation}
To obtain a stable cavity mode, we require that $q_1=q_2$ after one full round-trip. In a ring cavity, the non-normal incidence angles at the curved optics requires different ABCD matrices to treat the tangential plane separately from the sagittal plane.

\begin{figure}[tb]
\centering\includegraphics[width=\textwidth]{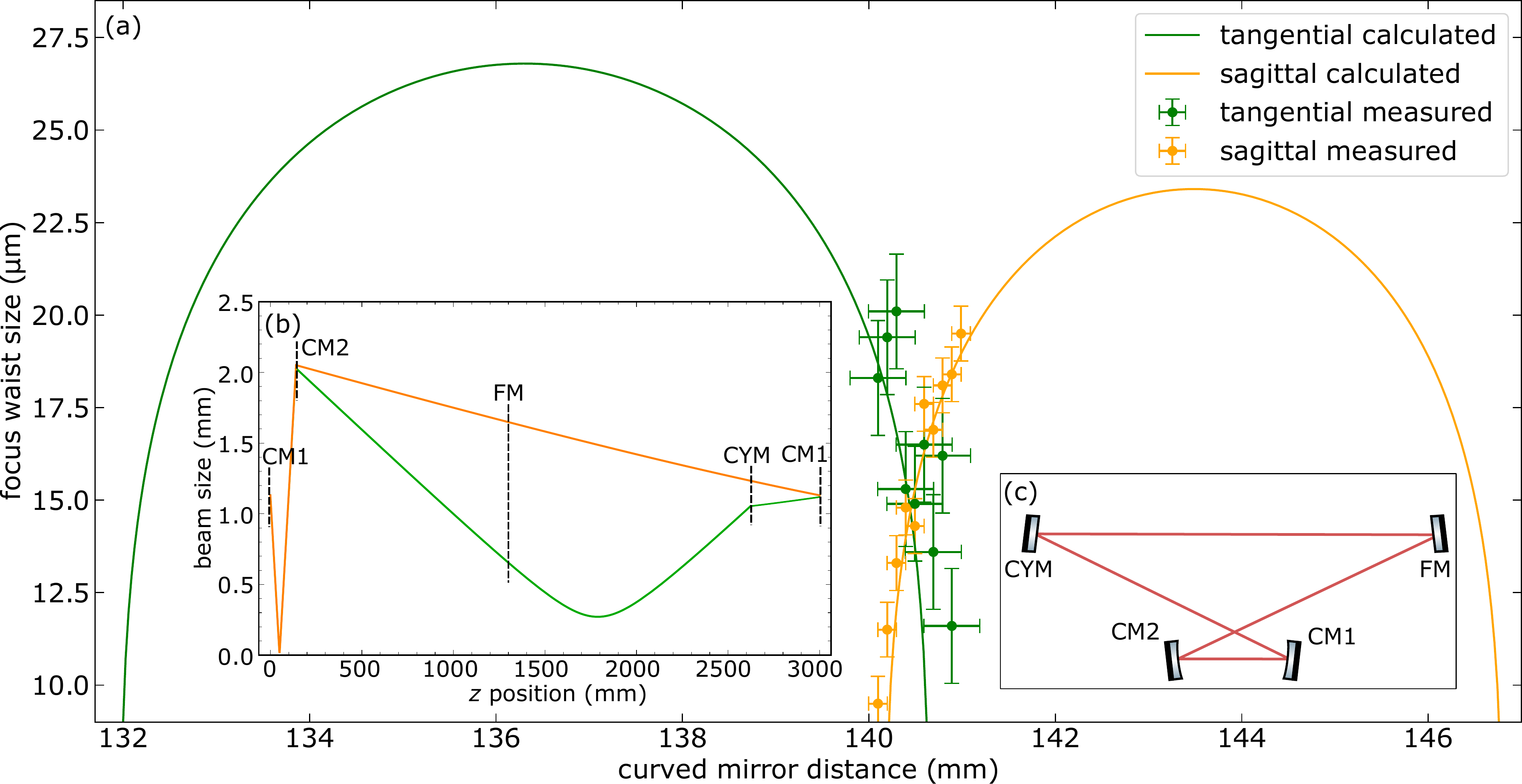}
\caption{(a) Focus waist size of the final cavity design. Due to the large incidence angles of 11\textdegree\ and 8\textdegree\ on the curved mirrors with ROC = 100\,mm and ROC = 175\,mm respectively, both stability regions are pulled apart, and only their small overlap remains for cavity operation. In its center, an astigmatism-free focus appears. Measurements confirmed stable cavity operation with a waist of 14.7\,$\mu$m in both planes. Inset (c) shows the overall geometry. (b) Calculated waist size of the cavity beam, with indicated mirror positions. A secondary tangential focus appears in the long cavity arm.}
\label{fig:fwrealcav}
\end{figure}


\section{Experimental setup}
Chirped pulses ($\sim$24\,ps) with 12\,W of average power and a 15\,nm FWHM bandwidth centered at 1039\,nm are produced by a 100\,MHz mode-locked laser (FC1000-250, Menlo Systems) referenced to the global positioning system signal. The pulses are amplified by a home-built amplifier consisting of an 80\,cm long Yb-doped photonic crystal-fiber rod (aeroGAIN-ROD-MODULE-2.0 PM85, NKT photonics) which is pumped backwards by a fiber-coupled 250\,W continuous wave laser diode (D4F2S22-976.3-250C-IS58.1, DILAS Diodenlaser GmbH) at 976\,nm. Using a 1000 lines/mm transmission grating (1158\_28x18\_6.35\_H, Gitterwerk GmbH) with a specified single-pass transmission $>98.5$\%, the pulse duration is compressed to 220\,fs at 88\,W of power. These pulses are then steered into an ultra-high vacuum chamber housing the fsEC. The optics for the 3-m long cavity are stably mounted on a rigid titanium rod structure that is directly seated on the optical table containing the FC. The surrounding vacuum chamber is mechanically decoupled from it by flexible bellows and resting on air pistons to avoid vibrations from the vacuum pumps reaching the mirrors. Details of the vacuum system as well as the multi-stage differential pump system surrounding the cavity focus will be described in a separate paper \cite{Nauta2020a}.

An overview of the in-vacuo optical elements is shown in Figure \ref{fig:cav_schem}(a). To monitor slow beam drifts, mirror leakage is directed to two CCD cameras outside the vacuum chamber. In this way, the alignment with the fsEC can recovered without opening the vacuum chamber when it is lost. Most of the light reflected from the IC is sent onto a beam dump, while a small part is spectrally dispersed and used for Pound-Drever-Hall-locking of the cavity length to the comb repetition rate. A 4\,MHz modulation signal is applied to an EOM inside the laser oscillator to produce an error signal, which is fed back onto a piezo-element behind the flat cavity mirror by a PID controller (STEMLAB 125-10 Redpitaya) \cite{Hannig2018}. For compensation of slow drifts of thermal origin, this mirror is mounted on a translation stage (Q-545.10U, Physik Instrumente) also controlled by the feedback loop, such that the fsEC can stay locked for many hours. 

A weak beam leaking from the cavity through the first curved mirror (CM1) is collimated and sent to a CCD camera (DCC1240M, Thorlabs) and a photodiode (PDA 100A, Thorlabs). The former continuously monitors the transverse profile of the cavity mode (see Figure \ref{fig:cav_schem}(d)); while the latter measures the intra-cavity power. We calibrate it by removing the input coupler and recording the photodiode voltage as function of incident power. For adjusting the distance between the two curved mirrors, CM1 is placed on a closed-loop translation stage (Q-521.24U, Physik Instrumente). Directly after the focus, the beam impinges on the GM (35\,mm x 20\,mm x 2\,mm substrate, Ibsen Photonics) under a 75\textdegree grazing-incidence angle. Its XUV diffraction efficiency for wavelengths between 40 and 100\,nm was simulated and optimized  using the software PCGrate. The resulting grating design (Figure \ref{fig:cav_schem}(b)) has a theoretical efficiency above 10\% in most of this range (Figure \ref{fig:cav_schem}(c)). The mounted GM has its backside glued to a 10\,mm$^2$ copper braid removing heat due to NIR absorption. To align the cavity in vacuum, the CYM and FPM are steered by piezoelectric picomotors (N-480 PiezoMike, Physik Instrumente). 

The cavity mirrors are coated by quarter-wave dielectric stacks (Layertec) of alternating SiO$_2$ and Ta$_2$O$_5$ layers. They generally ensure low dispersion and high damage thresholds at a reflectivity of $R \approx0.99996$ and a group-delay dispersion (GDD) $<$ 0.5\,fs$^2$ between 1020 and 1060\,nm. To set a low finesse, an IC mirror with a  reflectivity of $R_{\mathrm{IC}} = 0.993$ was chosen such that it dominates the cavity losses. The GM is designed to have a very high NIR reflectivity (R = 0.99999) with a GDD $<$ 2\,fs$^2$. However, the grating structure etched into its top layer causes some additional losses. The pulse-to-pulse phase accumulation inside the cavity is matched to the carrier-envelope offset (CEO) phase from the oscillator for maximum enhancement. Every few hours, we manually adjust the laser CEO frequency, which is measured by an f-2f interferometer. 

Harmonics angularly dispersed by the GM impinge on a plate coated with sodium salicylate, which under XUV irradiation fluoresces around 420\,nm. The spots from the harmonics are imaged with an electron-multiplying CCD camera (Luca R EMCCD, Andor), and the power of each one is measured with a GaAsP photodiode (G1127-04, Hamamatsu Photonics). Its sensitivity steeply drops above 680\,nm, making it insensitive to residual NIR radiation from the fsEC. The photodiode current is measured with a picoamperemeter (Model 152, Keithly), and normalized using a responsitivity curve provided by the Physikalisch-Technische Bundesanstalt in Braunschweig which was cross-checked with a power measurement using a 235\,nm laser.

\begin{figure}[tb]
\centering\includegraphics[width=\textwidth]{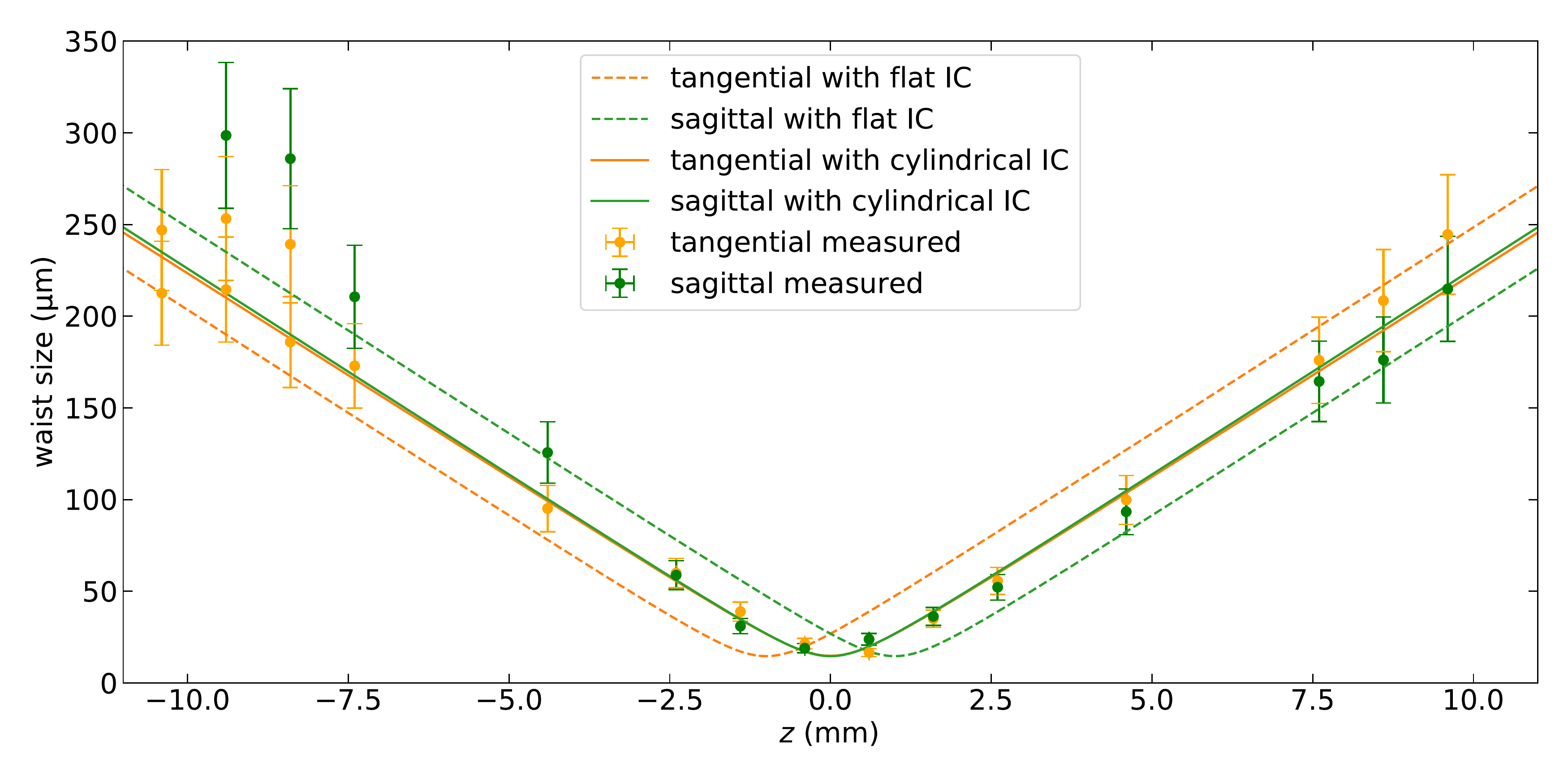}
\caption{Tangential and sagittal waist size as function of longitudinal position in the vicinity of the cavity focus. Without compensation, both foci are displaced by 2\,mm. By utilizing a cylindrical input coupler (IC), the foci can be made to overlap and produce a single, astigmatism-free focus. The effect of the cylindrical IC is confirmed by knife-edge measurements in both planes.}
\label{fig:focusposition}
\end{figure}


\section{Results \& discussion}
Providig sufficient space around the focus while keeping a small waist size results in large incidence angles on the curved cavity mirrors. This generates astigmatism and a separation of the vertical and horizontal foci, but at carefully chosen angles, tight foci can be produced in both the sagittal and tangential plane. Figure \ref{fig:fwrealcav}(a), shows the calculated focus waist sizes as a function of the curved mirror distance. The two stability regions only have a small overlap due to incidence angles of 11\textdegree\ and 8\textdegree\ on the curved mirrors with R = 175\,mm and R = 100\,mm, respectively. Yet, when the cavity is operated at the intersection of both regions, an astigmatism-free and tight focus is produced. In Figure \ref{fig:fwrealcav}(b), the resulting beam size throughout the cavity is shown. The secondary tangential focus in the long cavity arm is not problematic since its position is not close to a cavity mirror. 

To verify that the focus waist size of the implemented cavity corresponds to the calculations, the beam size of the transmitted cavity light was recorded on the CCD camera at two different positions in the beam. By back-propagating the Gaussian beam parameters, the focus waist size can be uniquely determined in each plane, as shown in Figure \ref{fig:fwrealcav}(a). Waist-size uncertainties arise from shifts of the optical elements in the transmitted beam path, and those of the curved-mirror distance from measurement errors. Both uncertainties are larger for the tangential than for the saggital plane, since the cavity had to be re-aligned in that plane during the measurement series because of the finite incidence angle on the curved mirrors. A good overall agreement between the measured and calculated waist sizes is found, demonstrating a focus waist of 14.7\,$\mu$m in both planes and stable cavity operation within the small overlap of both stability regions.

\begin{figure}[tb]
\centering\includegraphics[width=\textwidth]{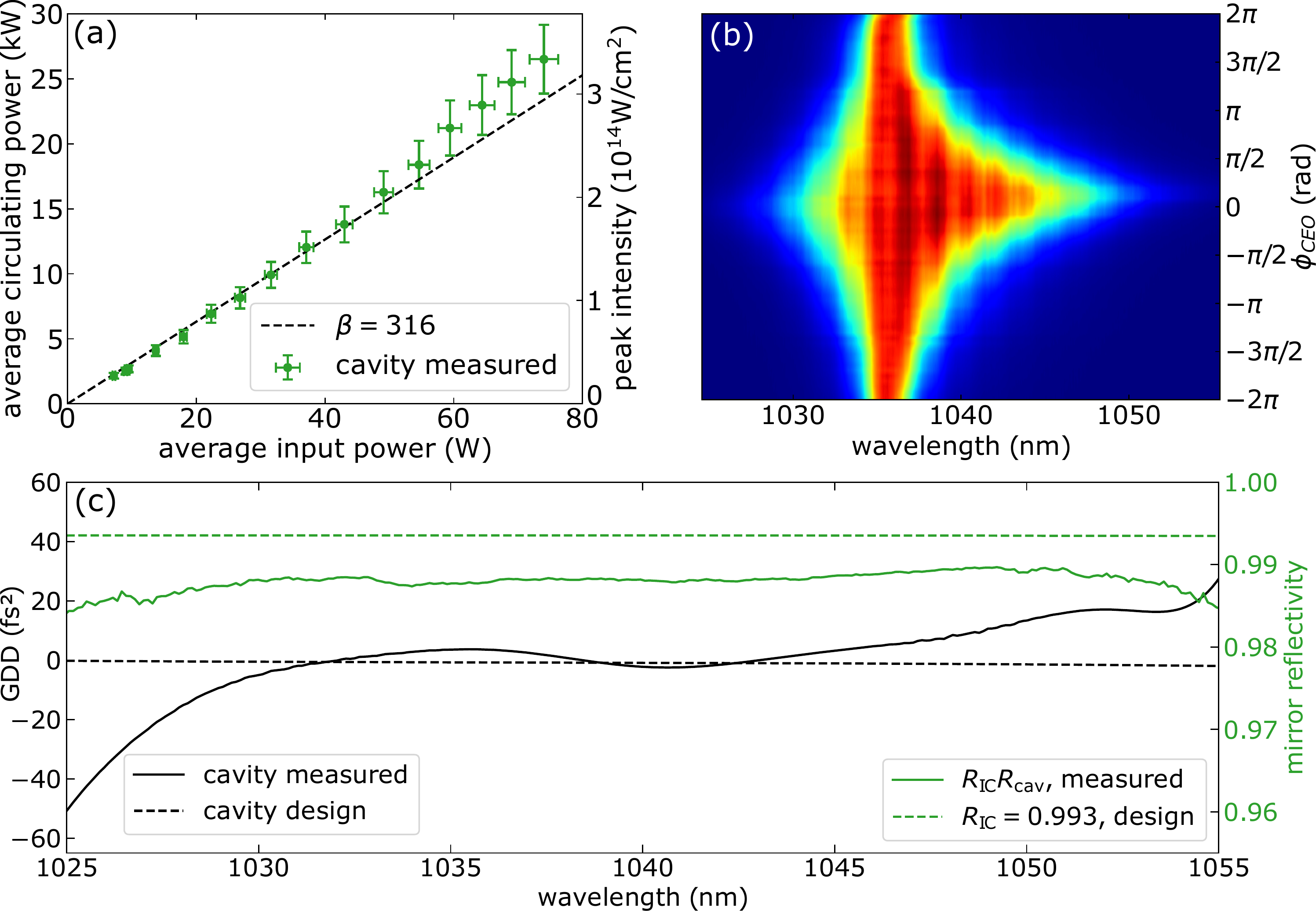}
\caption{(a) Intra-cavity power as function of incident power. With a $R_i=0.9935$ input coupler, an average enhancement of $\beta=316\pm8$ is reached for an empty cavity. An apparent super-linear behaviour is likely due to an increased sensitivity to calibration errors at very high circulating powers. (b) Color-coded cavity transmitted spectrum as function of carrier envelope offset phase $\phi_{\mathrm{CEO}}=2\pi\cdot f_{\mathrm{CEO}}/f_{\mathrm{rep}}$. The cavity was locked at a wavelength of 1035.6 nm during the measurement making this spectral range resonant for any phase offset. (c) Measured cavity GDD and total cavity reflectivity $R_{\mathrm{IC}} R_{\mathrm{cav}}$ as function of wavelength. The discrepancy between measured and theoretical values is most probably due to a damaged coating on one of the mirrors.}
\label{fig:cav_meas}
\end{figure}

Figure \ref{fig:focusposition} shows the waist size in longitudinal ($z$) direction; dashed lines indicate that the tangential and sagittal foci are displaced. This effect can be compensated by inserting a cylindrical IC with a custom curvature of 2061\,mm, as indicated by the solid lines. To corroborate this, we carried out knife-edge measurements of the beam size in the focus region. A razor blade mounted on an XYZ-translation stage was moved into the cavity beam while recording the transmitted intensity through one of the cavity mirrors on a photodiode. By fitting error functions to the intensity decay curves, we determine the beam waist data points in Figure \ref{fig:focusposition}. Large uncertainties are due to the interplay of edge-diffraction with the cavity mode \cite{Lee2019}. Data points near the focus confirm that full astigmatism compensation is accomplished.

\begin{figure}[tb]
\centering\includegraphics[width=.8\textwidth]{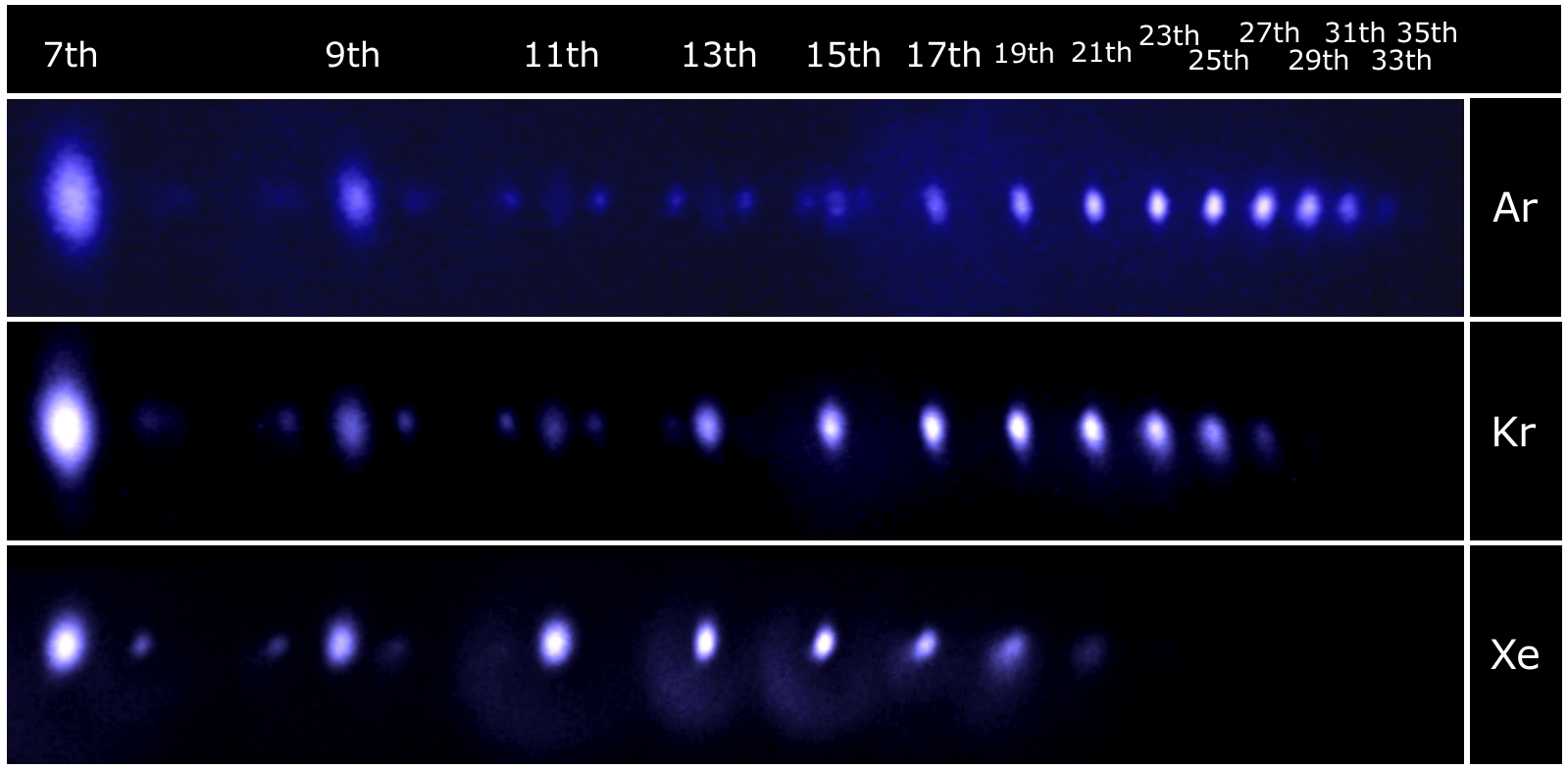}
\caption{Images of a screen coated with sodium salicylate showing fluorescence from various harmonic orders for three different target gases. Faint dots located between the odd harmonic orders originate from second-order diffraction from the grating mirror. The intensity scales for each gas are different.}
\label{fig:harm_screen}
\end{figure}

\begin{figure}[tb]
\centering\includegraphics[width=\textwidth]{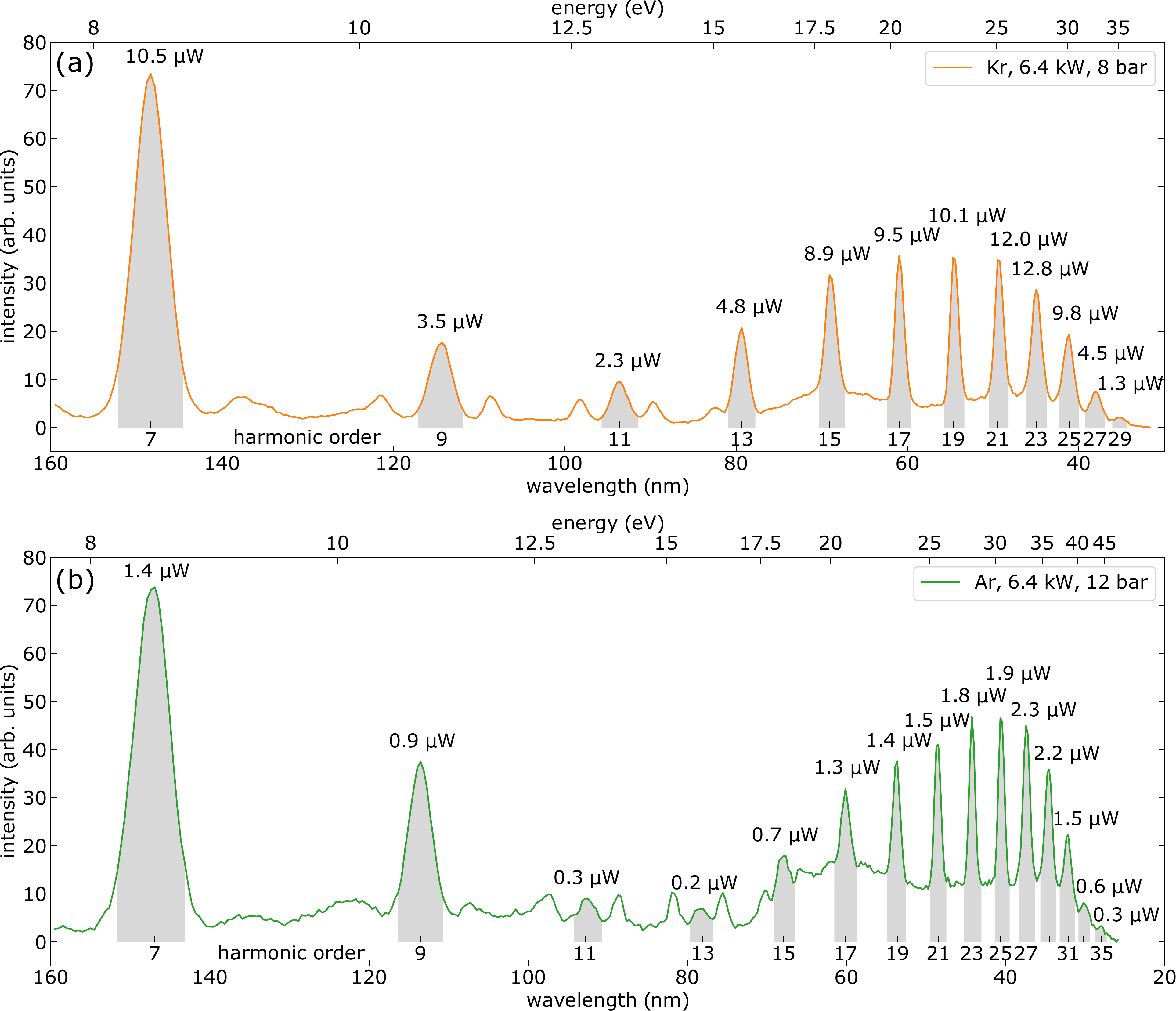}
\caption{Integrated harmonic yield at 6.4\,kW intra-cavity power for (a) Kr at 8\,bar; (b) Ar at 12 bar backing pressure. The power of the harmonics is measured with a photodiode normalized at the 15th and 17th harmonic, respectively.}
\label{fig:harm_yield}
\end{figure}

Figure \ref{fig:cav_meas}(a) shows the circulating power as a function of the incident power while sweeping the cavity length over the resonances. The fit yields an average enhancement factor of $316\pm8$. Two sets of cylindrical lenses are used for matching the incident beam to the cavity mode. The spatial mode-matching factor $\epsilon=0.75$ was determined by measuring the cavity contrast, defined as the ratio between the decrease of the reflected intensity at resonance and the off-resonance intensity, for various IC mirrors with a known reflectivity. Using the relation $\beta=\epsilon(1-R_{\mathrm{IC}})\frac{\mathcal{F}^2}{\pi^2}$, we obtain a cavity finesse $\mathcal{F}$ of~805. At an input power of 75\,W we reach an average circulating power of more than 25\,kW. For the 220\,fs pulse duration and the 679\,$\mathrm{\mu}$m$^2$ focal area, this yields a peak intensity at the focus of $3.3\times 10^{14}$\,W/cm$^2$, sufficient to drive the HHG process. 

By recording the cavity transmission spectrum while changing the comb offset frequency, the 2D color map shown in Figure \ref{fig:cav_meas}(b) was obtained. Its intensity depends on both the driving optical frequency $\omega$ and the comb CEO phase $\phi_{\mathrm{CEO}}$ and is proportional to the square of the intra-cavity electric-field amplitude
\begin{equation}
I_{\mathrm{trans}}\left(\omega,\phi_{\mathrm{CEO}}\right)\propto\left|E_{\mathrm{circ}}\left(\omega,\phi_{\mathrm{CEO}}\right)\right|^2=\left|\frac{\sqrt{1-R_{\mathrm{IC}}(\omega)}E_{inc}(\omega)}{1-\sqrt{R_{\mathrm{IC}}(\omega)R_{\mathrm{cav}}(\omega)}\exp\left[i\Delta \phi_{\mathrm{cav}}(\omega,\phi_{\mathrm{CEO}})\right]}\right|^2,
\label{eq:ceofit}
\end{equation}
where $R_{\mathrm{IC}}(\omega)$ is the IC reflectivity, $R_{\mathrm{cav}}(\omega)$ the product of all other cavity mirror reflectivities, and $\Delta \phi_{\mathrm{cav}}(\omega,\phi_{\mathrm{CEO}})$ the round-trip phase shift. For each vertical slice through the 2D intensity distribution, a value of $\Delta \phi_{\mathrm{cav}}$ can be determined for that specific wavelength, from which the cavity GDD can be extracted \cite{Schliesser2006}. The width of this peak provides a measure for the round-trip losses $\left(1-R_{\mathrm{cav}}\right)$. Fits were performed for many different wavelengths, yielding a mostly good agreement between the above model and the data. In Figure \ref{fig:cav_meas}(c), the resulting GDD values are shown (solid black line). Fast fluctuations in the calculated phase function are amplified when taking the numerical derivative in order to calculate the GDD, therefore a Gaussian convolution filter (7\,Thz standard deviation) was used to smooth the data before differentiating. The resulting GDD is still much larger than the calculated design values based on the technical specifications of the mirrors (dashed black line), especially at the wings of the spectrum. This could be due to the reduced signal-to-noise ratio there, or to the narrow bandwidth of our FC spectrum of 15\,nm making the measurement less accurate. Also, the large incidence angles of the curved mirrors could cause an increased dispersion in the wings of the spectrum, since the displayed design values are for normal incidence. 

The measured total cavity reflectivity $R_{\mathrm{IC}}R_{\mathrm{cav}}$ is displayed in Figure \ref{fig:cav_meas}(c) (solid green line) together with the theoretical reflectivity of the input coupler (dashed green line). As expected from our mirror design, the reflectivity is rather flat over the range of the incident spectrum, apart for some deviations in the wings. The difference between the measured curve and the IC reflectivity comes from additional losses inside the cavity $R_{\mathrm{cav}}$, which can be attributed largely to damaged spots on one mirror found after the measurements were taken. The overall flat cavity-loss spectrum proves that the mirror bandwidth is suitable to enhance the whole comb spectrum.

Figure \ref{fig:harm_screen} displays images of the fluorescence screen irradiated by harmonics generated with Ar, Kr and Xe as target gas. Their orders were identified by their position. Small dots between the uneven orders are due to second-order diffraction of higher-order harmonics. In the case of Xe, they appear at positions predicted for the 15th, 17th and 19th harmonic orders. Since the 21st and higher orders are much weaker, no second-order dots appear for those. Small, bright spots originate from phase-coherent short trajectories in the HHG process. Some harmonic orders display a fainter, larger halo, as the 13th and 15th in Xe or the 17th and 19th harmonic in Ar, which are caused by the long trajectories of the HHG process \cite{Gaarde1999}. In Xe we observe up to the 23th harmonic, for Kr up to the 29th and for Ar up to the 35th, corresponding to an energy (wavelength) of 42\,eV (30\,nm). The grating efficiency and the fluorescence yield of sodium salicylate steeply drop at such small wavelengths, limiting the highest observable harmonic order. In Figure \ref{fig:harm_yield}, the vertically integrated intensity of the fluorescent screen is shown for (a) Kr and (b) Ar at backing pressures of 8\,bar and 12\,bar, respectively, and a nozzle diameter of 50\,$\mathrm{\mu}$m. Due to its higher ionization potential, the phase-matching pressure for Ar is higher than that of Kr. For measuring the output power, each harmonic is integrated over the gray area, and the resulting total intensity normalized using the calibrated photodiode with an estimated uncertainty of 20\%. Deviations could be due to errors in the integration limits of the individual harmonics and to inhomogeneities in the thickness of the layer of sodium salicylate on the glass plate. 

\section{Conclusion}
We have presented a novel design for a fully astigmatism-compensated femtosecond cavity that enhances 100\,MHz, 80\,W, 220\,fs pulses from a phase-stabilized frequency comb. By introducing a cylindrical input coupler (IC) mirror, a round cavity focus with a waist size of 15\,$\mathrm{\mu}$m is produced with full symmetry in both the sagittal and tangential planes. The cavity enhances the complete incident spectrum by a factor of $316\pm8$, leading to focus intensities of $\sim 10^{14}$\,W/cm$^2$ sufficient for HHG. Produced XUV light is coupled out of the cavity using a grating mirror, angularly dispersing the harmonics for future frequency-resolved spectroscopy purposes. Space freed around the focus by our design accommodates a differential pump system needed for reaching high target gas densities in the focal volume while keeping ultra-high vacuum conditions in the chamber. Three different target gases, Xe, Kr and Ar were used, generating harmonics ranging from the 7th up to the 23th, 29th and 35th respectively, corresponding to a maximum observed photon energy (wavelength) of 42\,eV (30\,nm). 

Realizing this XUV comb is an important step towards performing the first ultra-high precision spectroscopy on HCI in this spectral region. Furthermore, the novel cavity enables multi-photon ionization experiments at a much lower intensity and higher precision than hitherto possible \cite{Nauta2020}. Moreover, the XUV comb could be employed to search for the nuclear transition in the $^{229m}$Th isomer, since the 7th harmonic of our comb matches with its most accurately determined energy \cite{Seiferle2019a}. To further improve the phase matching conditions and raise the HHG output, the setup is ready to use higher pressures and gas mixtures boosting the speed of the target gas and reducing the amount of detrimental steady-state plasma in the focus region \cite{Porat2018, Nauta2020a}.

\section*{Acknowledgments}
We thank the MPIK mechanical workshop for fabrication of numerous parts.

\bibliography{cavity_paper_bib_arxiv}
\end{document}